\documentclass[preprint,superscriptaddress,preprintnumbers,amsmath,amssymb,prd]{revtex4}
\usepackage{graphicx}

\begin{document}

\title{The Casimir pressure between metallic plates out of thermal
equilibrium: Proposed test for the relaxation properties of free
electrons}

\author{
G.~L.~Klimchitskaya}
\affiliation{Central Astronomical Observatory at Pulkovo of the
Russian Academy of Sciences, Saint Petersburg,
196140, Russia}
\affiliation{Institute of Physics, Nanotechnology and
Telecommunications, Peter the Great Saint Petersburg
Polytechnic University, Saint Petersburg, 195251, Russia}

\author{
V.~M.~Mostepanenko}
\affiliation{Central Astronomical Observatory at Pulkovo of the
Russian Academy of Sciences, Saint Petersburg,
196140, Russia}
\affiliation{Institute of Physics, Nanotechnology and
Telecommunications, Peter the Great Saint Petersburg
Polytechnic University, Saint Petersburg, 195251, Russia}
\affiliation{Kazan Federal University, Kazan, 420008, Russia}

\author{
R.~I.~P.~Sedmik }
\affiliation{Technische Universit\"{a}t Wien, Atominstitut, Stadionallee 2,
1020 Vienna, Austria}

\begin{abstract}
We propose a test on the role of relaxation properties of
conduction electrons in the Casimir pressure between two
parallel metal-coated plates kept at different temperatures.
It is shown that for sufficiently thick metallic coatings
the Casimir pressure and pressure gradient are determined
by the mean of the equilibrium contributions calculated
at temperatures of the two plates and by the term
independent on separation. Numerical computations of
the nonequilibrium pressures are performed for
two parallel Au plates of finite
thickness as a function of separation and temperature
of one of the plates using the plasma and Drude models
for extrapolation of the optical data of Au to low
frequencies. The obtained results essentially depend
on the extrapolation used. Modifications of the CANNEX
setup, originally developed to measure the Casimir
pressure and pressure gradient in thermal equilibrium,
are suggested, which allow different temperatures of
one of the plates. Computations of the nonequilibrium
pressure and pressure gradient are performed
for a realistic experimental configuration. According
to our results, even with only a 10~K difference in
temperature between the plates, the experiment could
discriminate between different theoretical
predictions for the total pressure and its
gradient, as well as for the contributions to them
due to nonequilibrium, at high confidence.
\end{abstract}

\maketitle
\newcommand{\ve}{\varepsilon}
\newcommand{\om}{\omega}

\section{Introduction}

Physical phenomena caused by quantum fluctuations of the electromagnetic
field attract much attention in both fundamental physics and its
applications \cite{1,2}. One of the most striking macroscopic effects of
this kind is the Casimir force \cite{3} resulting from the zero-point
and thermal fluctuations. This force manifests itself in many branches of
physics ranging from atomic physics, condensed matter physics to
elementary particle physics, gravitation and cosmology \cite{4}, and
is actively considered to be used in the next generation of microdevices
with reduced dimensions \cite{5,6,7,8,9,10,11}.

The theory of the Casimir force was developed by Lifshitz \cite{12} for the case
of two thick parallel plates (semispaces) of equal temperature in thermal
equilibrium with an environment. In the framework of this theory the
Casimir free energy of a fluctuating field and the force per unit area of
the plates (i.e., the Casimir pressure) are expressed via the
frequency-dependent dielectric permittivities of plate materials \cite{4,12}.
Over a long period of time, the comparison between experiment and theory
remained solely qualitative. Only during the last 15 years sufficiently
precise measurements have been performed which allow reliable quantitative
comparison between the measurement data and theoretical predictions
\cite{4,13}. The results of this comparison are commonly considered
as puzzling.
It turned out that the Lifshitz theory is in agreement with the experimental
data for metallic test bodies only under the condition that the relaxation
properties of free electrons are ignored in calculations. This was confirmed
by several experiments performed in two experimental groups using quite
different laboratory setups (see Refs.~\cite{4,13} for a review and
Refs.~\cite{13,14,15,16,17,18} for more recent results).
As a practical matter, this means that the low-frequency response of a metal
to the fluctuating electromagnetic field should be described by the lossless
dielectric permittivity of the plasma model rather than by the permittivity
of the lossy Drude model, which correctly describes the reaction of metals to
conventional (real) fields. Moreover, for metallic plates with perfect
crystal lattices the Lifshitz theory was found to be in agreement with
thermodynamics only when using the plasma model, and to violate the third
law of the thermodynamics (the Nernst heat theorem) when the Drude model
is used \cite{4,13,18a,18b,18c}.

Another phenomenon caused by quantum fluctuations is the radiative heat
transfer between two metallic bodies at different temperatures
\cite{19,20,21,22,23}. The first attack to the problem of generalized
Casimir force acting between two media varying in temperature and
separated by a gap was undertaken in Ref.~\cite{24}. The Casimir-Polder
atom-plate force and the Casimir force between metallic plates out of
thermal equilibrium were studied in Refs.~\cite{25,26}.
The complete theory of the Casimir interaction out of thermal equilibrium,
covering the plate-plate, plate-rarefied body, and atom-plate configurations,
was developed in Ref.~\cite{27}. At a later time this theory was generalized
to the systems of two or more bodies of arbitrary shape kept at different
temperatures which may be also different from the temperature of the
environment \cite{28,29,30,31,32,33,34,35,36,37}.
The radiative heat transfer at nonequilibrium was also investigated in
connection with the van~der~Waals friction force between moving bodies
\cite{38}. Apart from the semiclassical Lifshitz theory, where the
electromagnetic field is quantized but the material of the test bodies
is described by a classical dielectric function, the Casimir effect out of
equilibrium was also examined using a stochastic equation for the density
of Brownian charges moving on the background of a uniform
electroneutralizing charge \cite{40}.

Most of the calculations considering radiative heat transfer and nonequilibrium
Casimir forces between metallic surfaces performed up to date assume that
the response of metal to a low-frequency fluctuating field is described by
the Drude dielectric function taking into account the relaxation properties
of conduction electrons. To our knowledge, there are only two papers \cite{40a,41},
directed to the resolution of a puzzle in the comparison between theory and experiment
mentioned above, where the radiative heat transfer was calculated using
different types of response functions of a metal. It was found \cite{40a,41}
that the power of the heat transfer depends on the type of response
function, but available measurement data are not sufficient for making
convincing conclusions. According to the general theory developed in
Ref.~\cite{27}, the nonequilibrium Casimir force consists of three
contributions: the first one is expressed via the equilibrium Casimir forces
at two different temperatures, the second one is antisymmetric under the
interchange of temperatures (both of the two are separation-dependent), and
the third one which does not depend on separation between the plates.
In Ref.~\cite{42} a difference-force measurement was proposed which
allows reliable discrimination between theoretical predictions for the
antisymmetric contribution to the nonequilibrium Casimir force given by the
Drude and plasma models. This test requires the measurement of
small forces of the order of 1~fN, just as in Refs.~\cite{18,43}, and
it is not realized experimentally so far
(note that difference force measurement of the equilibrium Casimir forces
at different temperatures was suggested in Ref.~\cite{46a}).

In this paper, we propose another experimental possibility to
discriminate between two different theoretical approaches in the
nonequilibrium Casimir force using a modified setup of the CANNEX test
of the quantum vacuum (Casimir And Non-Newtonian force EXperiment)
\cite{44,45,46}. We calculate the nonequilibrium Casimir pressure
in the configuration of two parallel plates, each of which consists of
a sufficiently thick Au layer deposited on a dielectric substrate.
The thicknesses of the Au layers are chosen to be larger than the
characteristic wavelengths contributing to the thermal effect in the
Casimir force. It is assumed also that the upper plate is kept at the
laboratory temperature, whereas the temperature of the lower
plate is varied through some range.
In this case, the contribution to the Casimir pressure, which is antisymmetric
under the interchange of the plate temperatures, is equal to zero to a high
accuracy. As a result, only two other contributions, which have been screened
out in the setup of Ref.~\cite{42}, determine the total result.
Thus, our proposal is alternative to that one of Ref.~\cite{42}.
What is more, it relates to separations from 4 to $10~\mu$m,
whereas the differential measurements of Ref.~\cite{42} should be
performed at separations below $1~\mu$m.

Computations of the Casimir pressure at nonequilibrium conditions are
made using the optical data for Au extrapolated down to zero frequency
either by the Drude or by the plasma model for
the plates of finite thickness. The pressure is found as a function
of separation between the plates and of the temperature of the lower plate.
It is shown that the computational results obtained using the Drude
and plasma extrapolations can easily be discriminated over
wide ranges of separations and temperatures.

Additional computations of both the Casimir pressure and its gradient
are performed in the modified
configuration of the CANNEX test of the quantum
vacuum where, without sacrifice of precision, the temperature of
the lower plate can be increased by up to 10~K with respect to the
temperature of the upper plate.
It is shown that even at such a small temperature difference the theoretical
predictions using the Drude and plasma model extrapolations can be
reliably discriminated over the separation range from 4 to $10~\mu$m.
In so doing, measurements of the pressure gradient provide a test for
the first contribution to the nonequilibrium Casimir interaction
expressed via the equilibrium terms at two different temperatures,
whereas measurements of the pressure suggest a test for the additional
terms independent of separation.

The paper is organized as follows. In Sec.~II we summarize main results for
the out-of-equilibrium Casimir pressure in the form convenient for
computations. Section~III contains computational results for the
nonequilibrium Casimir pressure in the system of two Au plates
obtained using the Drude and plasma model extrapolations
of the optical data. In Sec.~IV we briefly describe the modified CANNEX
setup and perform computations of the nonequilibrium Casimir pressure and
pressure gradient in the experimental configuration. Section~V contains
our conclusions and discussion.

\section{The Casimir pressure between two parallel plates kept at
different temperatures}

Keeping in mind further applications of the CANNEX setup,
we consider the upper plate (1) as consisting of a thick dielectric substrate
having the dielectric permittivity $\ve_1(\om)$ and a metallic layer of
thickness $d_1$ having the dielectric permittivity $\ve_m(\om)$ deposited
on the lower surface. The temperature of the upper plate is $T_1$.
 In a similar way,
the lower plate (2) consists of a thick dielectric substrate
having the dielectric permittivity $\ve_2(\om)$ coated by a metallic layer of
thickness $d_2$ with dielectric permittivity $\ve_m(\om)$ deposited
on the top of it. The lower plate is kept at the temperature $T_2$.
It is separated by a distance $a$ from the upper one.

For this case
the Casimir pressure acting on the inside faces of the plates can be 
found in Refs.~\cite{27,31} (here we use the negative
sign for attractive pressures)
\begin{eqnarray}
&&
P(a,T_1,T_2)=\frac{1}{2}[P_{\rm eq}(a,T_1)+P_{\rm eq}(a,T_2)]
\nonumber \\
&&~~~~~~~
+\Delta P_{\rm neq}(a,T_1,T_2)+
\frac{2\sigma}{3c}(T_1^4+T_2^4),
\label{eq1}
\end{eqnarray}
\noindent
where $P_{\rm eq}$ is the standard equilibrium Casimir pressure at the respective temperature, $\Delta P_{\rm neq}$ is the term antisymmetric
under the interchange of temperatures, and $\sigma$ is the
Stefan-Boltzmann constant.

An explicit expression for $P_{\rm eq}$, convenient for numerical
computations, is given by \cite{4}
\begin{eqnarray}
&&P_{\rm eq}(a,T_j)=-\frac{k_BT_j}{8\pi a^3}
\sum_{l=0}^{\infty}{\vphantom{\sum}}^{\prime}
\int_{\zeta_l^{(j)}}^{\infty}y^2dy
\nonumber \\
&&~~~~~~
\times\sum_{\alpha}\left[\frac{e^y}{R_{\alpha}^{(1)}
(i\zeta_l^{(j)}\!\!,y)R_{\alpha}^{(2)}(i\zeta_l^{(j)}\!\!,y)}
-1\right]^{-1}\!.
\label{eq2}
\end{eqnarray}
\noindent
Here, $k_B$ is the Boltzmann constant, $j=1,\,2$, and the prime on the
summation sign divides the term with $l=0$ by 2. $\zeta_l^{(j)}$ with
$l=0,\,1,\,2,\,\ldots$ are the dimensionless Matsubara frequencies
connected with the dimensional ones by
\begin{equation}
\zeta_l^{(j)}=\frac{\xi_l^{(j)}}{\om_c}\equiv
\frac{2a\xi_l^{(j)}}{c}=
\frac{4\pi ak_BT_jl}{\hbar c}
\label{eq3}
\end{equation}
\noindent
and $y=2a(k_{\bot}^2+{\xi_l^{(j)}}^2/c^2)^{1/2}$, where $k_{\bot}$ is
the magnitude of the wave-vector projection on the plane of plates.

The reflection coefficients on the upper ($n=1$) and lower ($n=2$)
plates for two independent polarizations of the electromagnetic field,
transverse magnetic ($\alpha={\rm TM}$) and transverse electric
($\alpha={\rm TE}$), defined at the purely imaginary Matsubara frequencies
are given by
\begin{equation}
R_{\alpha}^{(n)}(i\zeta_l^{(j)}\!\!,y)=\frac{r_{\alpha}(1,\ve_{m,l}^{(j)})+
r_{\alpha}(\ve_{m,l}^{(j)},\ve_{n,l})
e^{-2d_nk(\ve_{m,l}^{(j)})}}{1+r_{\alpha}(1,\ve_{m,l}^{(j)})
r_{\alpha}(\ve_{m,l}^{(j)},\ve_{n,l})e^{-2d_nk(\ve_{m,l}^{(j)})}}.
\label{eq4}
\end{equation}
\noindent
Here, $\ve_{m,l}^{(j)}\equiv\ve_m(i\om_c\zeta_l^{(j)}),\quad
\ve_{n,l}\equiv\ve_n(i\om_c\zeta_l^{(j)})$, and
\begin{equation}
k(\ve)=\frac{1}{2a}\sqrt{(\ve-1){\zeta_l^{(j)}}^2+y^2}.
\label{eq5}
\end{equation}

The reflection coefficients on the boundary planes between vacuum and Au and
between Au and the dielectric substrates of the upper and lower plates
entering Eq.~({\ref{eq4}) are defined by
\begin{eqnarray}
&&
r_{\rm TM}(\ve,\tilde{\ve})=\frac{\tilde{\ve}k(\ve)-
\ve k(\tilde{\ve})}{\tilde{\ve}k(\ve)+\ve k(\tilde{\ve})},
\nonumber \\
&&
r_{\rm TE}(\ve,\tilde{\ve})=\frac{k(\ve)-
k(\tilde{\ve})}{k(\ve)+ k(\tilde{\ve})}.
\label{eq6}
\end{eqnarray}
\noindent
In so doing the dielectric substrates are assumed to be infinitely thick.
Note that with respect to the reflectivity properties of dielectric
substrates this assumption is valid
 if the substrate thickness is larger than
$2~\mu$m \cite{47}.
In the proposed experiment, the dielectric substrates are much thicker
(see Sec.~IVA). Furthermore, as shown in Sec.~IVB, with
the actual experimental
thicknesses of Au layers $d_1$ and $d_2$ the dielectric parts of the
plates do not contribute to the results [in situations when the finite
thickness of the dielectric substrates is essential for the calculation of
reflection coefficients, one should use the well known generalization
\cite{51T,51R} of Eq.~(\ref{eq6}) which has the same form as
 Eq.~(\ref{eq4})].

The term $\Delta P_{\rm neq}$ in Eq.~(\ref{eq1}) can be most conveniently
 expressed in terms of the dimensionless integration variables
 $u=\om/\om_c$ and $t=k_{\bot}c/\om$. Introducing these variables in the
 respective expressions of Refs.~\cite{27,31}, one obtains
\begin{eqnarray}
&&
\Delta P_{\rm neq}(a,T_1,T_2)=\frac{\hbar c}{64\pi^2a^4}
\int_{0}^{\infty}\!\!\!\!\!u^3du\left[n(u,T_1)-n(u,T_2)\right]
\nonumber \\
&&~~~~~~\times
 \sum_{\alpha}\left[
\int_0^1t\sqrt{1-t^2}dt\frac{\left|R_{\alpha}^{(2)}(u,t)\right|^2-
\left|R_{\alpha}^{(1)}(u,t)\right|^2}{|D_{\alpha}(u,t)|^2}
\right.
\label{eq7}\\
&&~~~~~~~
-2\int_1^{\infty}t\sqrt{t^2-1}e^{-u\sqrt{t^2-1}}dt
%\label{eq7} \\
%&&
\left.%\times
\frac{{\rm Im}R_{\alpha}^{(1)}(u,t){\rm Re}R_{\alpha}^{(2)}(u,t)-
{\rm Re}R_{\alpha}^{(1)}(u,t)
{\rm Im}R_{\alpha}^{(2)}(u,t)}{|D_{\alpha}(u,t)|^2}
\vphantom{\int_0^1\frac{\left|R_{\alpha}^{(2)}(u,t)\right|^2-
\left|R_{\alpha}^{(1)}(u,t)\right|^2}{|D_{\alpha}(u,t)|^2}}
\right].
\nonumber
\end{eqnarray}
\noindent
Here, we have introduced the following notations:
\begin{eqnarray}
&&
n(u,T_j)=\left[\exp\left(
\frac{\hbar\om_cu}{k_BT_j}\right)-1\right]^{-1},
\label{eq8} \\
&&
D_{\alpha}(u,t)=1-R_{\alpha}^{(1)}(u,t)R_{\alpha}^{(2)}(u,t)
\exp(iu\sqrt{1-t^2}).
\nonumber
\end{eqnarray}
\noindent
Note that the first term in the square brackets in Eq.~(\ref{eq7})
results from the propagating waves and contains the contribution
independent on $a$. The second
term results from the evanescent waves.
Note also that although Eq.~(\ref{eq7}) assumes temperature-independent 
dielectric permittivity of a metal $\ve_m$, it can be applied to metals 
described by the Drude model (see Sec.~III), where the temperature 
dependence of the relaxation parameter makes only a minor impact 
on the computational results \cite{42}.

The reflection coefficients on the upper and lower plates at real
frequencies in terms of the new variables $u$ and $t$ take the form
\begin{equation}
R_{\alpha}^{(j)}(u,t)=\frac{r_{\alpha}(1,\ve_{m})+
r_{\alpha}(\ve_{m},\ve_{j})
e^{2id_jk(\ve_{m})}}{1+r_{\alpha}(1,\ve_{m})
r_{\alpha}(\ve_{m},\ve_{j})e^{2id_jk(\ve_{m})}},
\label{eq9}
\end{equation}
\noindent
where $\ve_{m}\equiv\ve_m(\om)=\ve_m(\om_cu),\quad
\ve_{j}\equiv\ve_j(\om)=\ve_j(\om_cu)$ and
\begin{equation}
k(\ve)=\frac{u}{2a}\sqrt{\ve-t^2}.
\label{eq10}
\end{equation}
\noindent
The reflection coefficients $r_{\alpha}(1,\ve_{m})$ and
$r_{\alpha}(\ve_{m},\ve_{j})$ entering Eq.~(\ref{eq9}) are
again expressed via Eq.~(\ref{eq6}) where the quantity $k(\ve)$
is now defined by Eq.~(\ref{eq10}) and all dielectric permittivities
are taken along the real frequency axis as functions of $u$.

Laboratory test bodies are usually placed in an environment with
some temperature $T_3$.
 This results in additional external pressures on their outside surfaces
depending on the reflectivity properties of these surfaces and the
temperature of the environment. Assuming that the external surfaces
of dielectric substrates are blackened, the total pressure on plate $i$
is given by \cite{31}
\begin{equation}
P_{\rm tot}^{(i)}(a,T_1,T_2,T_3)=P(a,T_1,T_2)-
\frac{2\sigma}{3c}(T_i^4+T_3^4),
\label{eq10a}
\end{equation}
\noindent
where $P(a,T_1,T_2)$ is given by Eq.~(\ref{eq1}).

If the temperature of the environment is the same as
the one of the upper plate,
i.e., $T_3=T_1$,
one obtains
different total pressures acting on the upper
\begin{eqnarray}
&&
P_{\rm tot}^{(1)}(a,T_1,T_2)=P(a,T_1,T_2)-\frac{4\sigma}{3c}T_1^4
\nonumber\\
&&~~~~
=\frac{1}{2}[P_{\rm eq}(a,T_1)+P_{\rm eq}(a,T_2)]
\label{eq10a} \\
&&~~~~
+\Delta P_{\rm neq}(a,T_1,T_2)+
\frac{2\sigma}{3c}(T_2^4-T_1^4),
\nonumber
\end{eqnarray}
\noindent
and lower plates
\begin{eqnarray}
&&
P_{\rm tot}^{(2)}(a,T_1,T_2)=P(a,T_1,T_2)-\frac{2\sigma}{3c}(T_2^4+T_1^4)
\label{eq10b}\\
&&~~~~
=\frac{1}{2}[P_{\rm eq}(a,T_1)+P_{\rm eq}(a,T_2)]
+\Delta P_{\rm neq}(a,T_1,T_2).
\nonumber
\end{eqnarray}

In the next section, as a model example,
 we calculate the out-of-thermal equilibrium
Casimir pressure between two gold plates of equal thicknesses
$d_1=d_2$ kept at temperatures
$T_1$ and $T_2$ separated by a gap of width $a$. This simple
configuration can be described as a particular case of the above
formulas with $\ve_1=\ve_2=1$. As a consequence, from Eq.~(\ref{eq9})
one obtains
\begin{equation}
R_{\alpha}^{(1)}(u,t)=R_{\alpha}^{(2)}(u,t)
\label{eq11}
\end{equation}
\noindent
and from Eq.~(\ref{eq7}) arrives at $\Delta P_{\rm neq}(a,T_1,T_2)=0$.
Then Eqs.~(\ref{eq10a}) and (\ref{eq10b}) result in
\begin{eqnarray}
&&
P_{\rm tot}^{(1)}(a,T_1,T_2)=\frac{1}{2}[P_{\rm eq}(a,T_1)+P_{\rm eq}(a,T_2)]
\nonumber \\
&&~~~~~~~
+
\frac{2\sigma}{3c}(T_2^4-T_1^4),
\label{eq13}\\
&&
P_{\rm tot}^{(2)}(a,T_1,T_2)=\frac{1}{2}[P_{\rm eq}(a,T_1)+P_{\rm eq}(a,T_2)],
\label{eq12}
\end{eqnarray}
\noindent
i.e., the pressure of the lower plate is expressed exclusively in terms of
the equilibrium pressures at two different temperatures.
Note that  Eqs.~(\ref{eq13}) and (\ref{eq12}) are proven under the condition that
the upper plate is in thermal equilibrium with the environment
at temperature $T_1$, whereas the lower plate is not.

\section{Computational results for gold plates}

Here, we perform numerical computations of the  pressures
$P_{\rm tot}^{(1,2)}(a,T_1,T_2)$ acting on two Au plates  kept at
temperatures $T_1$ and  $T_2$ using Eqs.~(\ref{eq13}) and (\ref{eq12}).
For simplicity we take $d_1=d_2=1~\mu$m, i.e., much larger than the
penetration depths of electromagnetic fluctuations in Au. Then
the computational results do not depend on the thickness of
plates.
Extrapolations of the optical data of Au to zero frequency
are made by means of the Drude and the plasma models.
Both extrapolations were
used extensively in calculations of the Casimir force (see Sec.~I).

The most standard source of optical data is Ref.~\cite{48}.
An extrapolation of these data taking into account the relaxation of
conduction electrons under the influence of an external electromagnetic
field is made by means of the lossy Drude model
\begin{equation}
\ve_D(\om)=1-\frac{\om_p^2}{\om[\om+i\gamma(T)]},
\label{eq14}
\end{equation}
\noindent
where $\hbar\om_p=9.0~$eV is the plasma frequency and
$\hbar\gamma(T)=0.035~$eV is the relaxation parameter of Au
at $T=T_1=300~$K.

By putting $\gamma(T)=0$ in Eq.~(\ref{eq14}) one obtains the lossless
plasma model
\begin{equation}
\ve_p(\om)=1-\frac{\om_p^2}{\om^2},
\label{eq15}
\end{equation}
\noindent
which is commonly used in the frequency region of infrared optics where
$\om\gg\gamma(T)$ and the relaxation properties do not play any role.
As to the Casimir forces in thermal equilibrium, caused by the fluctuating
fields, it was found, however (see Sec.~I), that for agreement with experimental
data and with the principles of thermodynamics one should
use the extrapolation of the optical data down to zero frequency by means
of Eq.~(\ref{eq15}) rather than Eq.~(\ref{eq14}). Detailed information concerning both extrapolations can be found
in Refs.~\cite{4,13}.

Now we use the resulting dielectric permittivities
of Au along the imaginary frequency axis to calculate the Casimir
pressure out of thermal equilibrium.
We first consider the  pressure $P_{\rm tot}^{(2)}$ acting on the
lower Au plate kept at temperature $T_2$, which does not contain
the separation-independent term.
In order to clearly demonstrate the effect of thermal nonequilibrium, we choose
$T_1=300~$K (as in the environment) and a rather large $T_2=500~$K.
In Fig.~\ref{fg1} we plot the resulting magnitude of the (negative)
nonequilibrium Casimir pressure as a function of separation, computed with the plasma (top solid line) and Drude (bottom solid line) models for extrapolation of the
optical data of Au.
As a comparison, we also plot the corresponding results for the equilibrium case ($T_1=T_2=300~$K) as the dashed lines, again using the plasma (top line) and Drude
(bottom line) extrapolations, respectively.

As is seen in Fig.~\ref{fg1}, the magnitudes of the Casimir pressure
out of thermal equilibrium are larger than in equilibrium if the
plasma model extrapolation is used. If, however, the Drude model
extrapolation is used in computations, the bottom solid and dashed
lines intersect, i.e., at short separations the magnitude of the
nonequilibrium Casimir pressure is smaller than of the equilibrium
one. To demonstrate this effect more clearly, we show the relevant separation range 1.5--$3~\mu$m in the inset. It is seen that if the Drude model is used an intersection
between the solid and dashed lines takes place at $a\approx 2.3~\mu$m.
As can be seen in Fig.~\ref{fg1}, thermal nonequilibrium has a strong impact on the magnitude of the Casimir pressure.

Now we calculate the nonequilibrium Casimir pressure $P_{\rm tot}^{(2)}$
as a function of the temperature $T_2$ of the lower plate
with fixed $T_1=300~$K.
The computational results are normalized by the Casimir
pressure at equilibrium, $P_{\rm \!eq}$, computed at $T_1=T_2=300~$K
and are shown in Fig.~\ref{fg2}. The three pairs of lines labeled 1, 2,
and 3 in Fig.~\ref{fg2}(a) are plotted at separations $a=1$, 2, and
$2.5~\mu$m, respectively. In so doing, the top line in each pair is
computed using the plasma model extrapolation and the bottom one using the
Drude model extrapolation. In a similar way, in Fig.~\ref{fg2}(b) the two
pairs of lines labeled 1 and 2 are plotted at $a=3$ and $5~\mu$m, and the
top and bottom lines in each pair are computed using the plasma and Drude
models, respectively.

As is seen in Fig.~\ref{fg2}, the ratio $P_{\rm tot}^{(2)}/P_{\rm \!eq}$
monotonously increases with $T_2$ if the plasma model is used in
computations. If, however, the Drude model is used, the quantity
$P_{\rm tot}^{(2)}/P_{\rm \!eq}$ is nonmonotonous with increase of $T_2$
[see the bottom lines in Fig.~\ref{fg2}(a)]. At all separations and
temperatures there are significant differences between the computational
results obtained using the plasma and Drude extrapolations of the
optical data [note that although at $a=5~\mu$m the two solid lines labeled 2
in Fig.~\ref{fg2}(b) deviate by only 1--2\%, the theoretical predictions
of the Lifshitz theory combined with either the Drude or the plasma models
for both $P_{\rm tot}^{(2)}$ and $P_{\rm eq}$ differ by almost
a factor of two].

Next, we consider the pressure on the upper plate which is in equilibrium
with the environment at temperature $T_1$.
It is given by Eq.~(\ref{eq13}).
 We put $T_1=300~$K equal to the temperature of the environment and
$T_2=500~$K.
The computational results as a function of separation
are shown in Fig.~\ref{fg3} by the solid lines
labeled 1 and 2, which are computed using extrapolations of the optical
data by means of
the plasma and Drude models, respectively. For comparison purposes,
the dashed lines 1 and 2 show the magnitudes of the total pressure
on the lower plate computed by Eq.~(\ref{eq12}) using the same
respective extrapolations of the optical data
(these dashed lines were already presented as the respective solid
lines in Fig.~\ref{fg1}).
Note that for the upper plate the nonequilibrium
pressure changes its sign from attractive to repulsive. This happens due
to the presence of the last term on the right-hand side of Eq.~(\ref{eq13}).
Thus, for each of the solid lines 1 and 2 in Fig.~\ref{fg3} the range
of separations to the left of the respective minimum corresponds to the
attraction (the force is negative), and to the right of the respective
minimum corresponds to  repulsion (the force is positive).
The values of separation distances separating the ranges of attraction
and repulsion are $a\approx 4.3$ and $3.5~\mu$m when the plasma and the
Drude model extrapolations are used in computations, respectively.

\section{Experimental test capable of discriminating between different
theoretical approaches}

In this section we suggest minor modifications in the experimental setup
of the CANNEX test of the quantum vacuum, which was originally suggested for
observation of  thermal effects in the equilibrium Casimir force
at large separations and constraining Yukawa-type corrections to
Newton's gravitational law and parameters of hypothetical particles
\cite{44,45,46}. We demonstrate that with these modifications CANNEX
is most useful for a measurement of the nonequilibrium
pressure considered in Secs.\ II and III and for a conclusive
discrimination between different theoretical approaches to the
account of free charge carriers.

\subsection{The modified CANNEX setup}

CANNEX is an experimental setup for simultaneous
measurements of the Casimir pressure and its gradient on the upper
one of two parallel plates at separations from 3 or 4 to $15~\mu$m.
The schematic of the setup is presented in Fig.~\ref{fg4}.
The upper (sensor) plate is a disc of $R=5.742$~mm radius
consisting of a Si substrate of $100~\mu$m thickness and an Au
film of thickness $d=200$~nm deposited on its bottom surface.
The system of a sensor plate attached to three springs (only two of them
are shown symbolically) is
characterized by the elastic constant $k$ and effective mass
$m_{\rm eff}$, and has the resonance frequency
$\om_0=(k/m_{\rm eff})^{1/2}$ \cite{44}.
The lower plate is formed by a vertical SiO$_2$ cylinder of
6~mm height coated with $1~\mu$m of Au.
The entire setup is placed in a vacuum chamber (see
Refs.~\cite{45,46} for more details).

The pressure applied to the upper plate, as well as its gradient, are detected
interferometrically in the experiment.
Pressures are sensed by monitoring changes in the extension of the sensor's
springs, $\Delta a=\pi R^2P/k$.
Pressure gradients are measured using a phase-locked loop that
detects the shift $\Delta\om$ of the resonance frequency under
the influence of the total (Casimir)
force \cite{4,13,14,45}. For this purpose, the lower three interferometers
(again, only two of them are shown
in the simplified scheme of Fig.~\ref{fg4}) are used. The latter also monitor
the separation distance
$a$ at different positions around the rim of the lower plate,
thereby allowing for an accurate determination and control of parallelism.
As a result of several improvements in the setup suggested
in Ref.~\cite{46}, sensitivities of the setup relative to the
pressure and pressure gradient measurements will be improved
to 1~nPa and 1~mPa/m, respectively.

In addition to the recent proposal~\cite{46}, we now present another modification of the CANNEX setup, allowing to increase the
temperature of the lower plate by 10~K with no
loss in the sensitivity. In fact, the improved sensitivities
mentioned above require a temperature stability of the sensor better than $1\,$mK.
For this reason, only the temperature of the lower plate may be varied,
while the sensor plate has to be kept stable in temperature.
This can practically be achieved by the thermal measurement and control scheme sketched in Fig.~\ref{fg4}.
The upper half of the sensor plate is connected radiatively to a thermal
shroud mounted on a Peltier element. In order to monitor the sensor's
 temperature, a contactless thermopile sensor is placed above the upper plate.
On the lower side, the temperature of the fixed SiO$_2$ plate is measured
near its surface via an embedded platinum resistor, and controlled by
Peltier elements at its base.

At a controlled temperature of $T_1=300\,$K of the sensor plate
and entire setup (which is the temperature of an environment),
and a temperature $T_2=310~$K of the lower plate,
the net radiative input to the sensor plate is just $129~\mu$W, thanks to
the low thermal emission and absorption coefficient of gold ($\sim\, 0.02$).
The resulting temperature gradient over the thickness of the sensor is negligible.
Moreover, as Si is almost transparent at wavelengths larger
than the bandgap ($1.1~\mu$m) but has an emissivity of around
0.7~\cite{49}, the thermal shroud can be in perfect thermal contact with
the sensor. Keeping the sensor at $T=T_1$ thus requires the shroud
temperature to be roughly 301~mK below $T_1$. Excess heat can always
be eliminated via heat pipes and radiators that interact with the inner
wall of the vacuum chamber (not shown), which is temperature controlled
with a precision of $1~$mK as well.

\subsection{Computational results in the experimental configuration}

Now we compute the nonequlibrium total pressure and pressure gradient
on the upper plate for the experimental parameters listed above including
the values of temperature $T_1=300~$K and $T_2=310~$K, i.e., a rather
moderate change, as compared to the equilibrium situation.
It turns out, however, that this change is quite sufficient in order
to observe the role of nonequilibrium in the measured quantities
as well as to discriminate between different theoretical approaches to
the description of relaxation using the CANNEX setup.

We start with the computation of the pressure gradient which is not sensitive
to the presence of the third (separation-independent) term on the
right-hand side of Eq.~(\ref{eq10a})
\begin{eqnarray}
&&
{P_{\rm tot}^{(1)}}^{\,\prime}(a,T_1,T_2)=\frac{1}{2}[P_{\rm eq}^{\,\prime}(a,T_1)+
P_{\rm eq}^{\,\prime}(a,T_2)]
\nonumber \\
&&~~~~~~~
+\Delta P_{\rm neq}^{\,\prime}(a,T_1,T_2).
\label{eq16}
\end{eqnarray}
\noindent
Direct computations using Eqs.~(\ref{eq7})--(\ref{eq10}) show that although
the parallel plates of CANNEX are dissimilar, with the experimental
parameters $d_1=200$~nm and $d_2=1~\mu$m the contribution of the term
$\Delta P_{\rm neq}^{\,\prime}$ on the right-hand side of Eq.~(\ref{eq16})
is by more than four orders of magnitude less than the contribution of
the first term. This is explained by the fact that the thicknesses
$d_1$ and $d_2$ are larger than the thermal wavelength contributing
to $\Delta P_{\rm neq}^{\,\prime}$. Thus, in the experimental configuration
the gradient of the total pressure
on the upper plate can be computed by the equation
\begin{equation}
{P_{\rm tot}^{(1)}}^{\,\prime}(a,T_1,T_2)=\frac{1}{2}[P_{\rm eq}^{\,\prime}(a,T_1)+
P_{\rm eq}^{\,\prime}(a,T_2)],
\label{eq16a}
\end{equation}
\noindent
where from Eq.~(\ref{eq2}) one obtains
\begin{eqnarray}
&&P_{\rm eq}^{\,\prime}(a,T_j)=\frac{k_BT_j}{8\pi a^4}
\sum_{l=0}^{\infty}{\vphantom{\sum}}^{\prime}
\int_{\zeta_l^{(j)}}^{\infty}y^3dy
\nonumber \\
&&~
\times\sum_{\alpha}\left[\frac{e^y}{R_{\alpha}^{(1)}
(i\zeta_l^{(j)}\!\!,y)R_{\alpha}^{(2)}(i\zeta_l^{(j)}\!\!,y)}
-1\right]^{-2}\frac{e^y}{R_{\alpha}^{(1)}
(i\zeta_l^{(j)}\!\!,y)R_{\alpha}^{(2)}(i\zeta_l^{(j)}\!\!,y)}.
\label{eq17}
\end{eqnarray}
\noindent
Here, the reflection coefficients are defined in Eqs.~(\ref{eq4})--(\ref{eq6}).

In Fig.~\ref{fg5}, we present computational results for the gradient of the
 total pressure applied to the upper plate for the experimental configuration.
The top and bottom lines are obtained when the extrapolation of optical data
of Au to low frequencies is performed by
means of the plasma and Drude models, respectively.
Note that with the experimental thicknesses $d_1$ and $d_2$ of the
Au layers indicated above, the dielectric parts of the plates
do not contribute to the result.
As is seen in Fig.~\ref{fg5},
within the separation region from 4 to $9~\mu$m the
differences in the two theoretical predictions exceed the experimental
sensitivity in measurements of the pressure gradient by a factor of
$2\times 10^3$ to $10^2$. This means that the
alternative theoretical approaches to the calculation of the
pressure gradient can be easily discriminated in the experiment
under consideration.

Now we discuss whether it is possible to discriminate the
contribution due to different temperatures of the plates in the
measurement results for the total nonequilibrium pressure
gradient ${P_{\rm tot}^{(1)}}^{\,\prime}(a,T_1,T_2)$ given by Eq.~(\ref{eq16a}).
For this purpose we consider the differential pressure gradient
\begin{equation}
{P_{\rm diff}^{(1)}}^{\,\prime}(a,T_1,T_2)={P_{\rm tot}^{(1)}}^{\,\prime}(a,T_1,T_2)-
P_{\rm eq}^{\,\prime}(a,T_1),
\label{eq18}
\end{equation}
\noindent
where $T_1=300~$K, $T_2=310~$K.

The computational results for the quantity ${P_{\rm diff}^{(1)}}^{\,\prime}$
are shown in Fig.~\ref{fg6} as functions of separation. The top
and bottom lines are computed using the plasma and Drude extrapolations
of the optical data of Au to low frequencies, respectively.
According to Sec.~IV.A, the experimental sensitivity with respect to
the difference of two pressure gradients is 2~mPa/m.
As is seen in Fig.~\ref{fg6}, the differential pressure gradients
shown by the bottom line computed using the Drude extrapolation
exceed the experimental sensitivity by up to a factor 4.
However, the differential pressure gradients
computed using the plasma extrapolation (the top line)
exceed the experimental sensitivity by up to a factor 18.
Thus, even for only 10~K temperature difference between the plates
both the effect of nonequlibrium and the type of theoretical approach
used for its description can be reliably determined.

Now we return to the nonequlibrium total pressure given by
Eq.~(\ref{eq10a}) and consider potentialities of CANNEX as a test
for the presence of separation-independent contributions.
First, we calculate the pressure applied to the upper plate kept at $T_1=300~$K
while the lower plate is kept at $T_2=310~$K. The computational
results as functions of separation are shown in Fig.~\ref{fg7}
where the top and bottom solid lines are computed using the Drude and plasma
extrapolations of the optical data to low frequencies, respectively.
Taking into account that the sensitivity of the CANNEX setup to pressure
measurements is equal to 1~nPa, the alternative theoretical predictions
can be easily discriminated experimentally over the entire separation
range from 4 to $10~\mu$m shown in Fig.~\ref{fg7}.
The dashed lines in Fig.~\ref{fg7} show the respective computational
results with omitted separation-independent term in Eq.~(\ref{eq10a}). The difference
between the solid and neighboring dashed lines is equal to $0.14~\mu$Pa
and, thus, can be observed in the CANNEX experiment.

As is seen in Fig.~\ref{fg7}, the
presence of a separation-independent term in
Eq.~(\ref{eq13}) does not lead to some
qualitative changes in the total  pressure.
To demonstrate the role of this term in more detail, we consider the differential
 pressure applied to an upper plate
\begin{equation}
P_{\rm diff}^{(1)}(a,T_1,T_2)=
P_{\rm tot}^{(1)}(a,T_1,T_2)-P_{\rm eq}(a,T_1).
\label{eq19}
\end{equation}

The computational results for the quantity $P_{\rm diff}^{(1)}$
are shown as functions of separation by the top
and bottom solid lines in Fig.~\ref{fg8}
obtained using the Drude and plasma models, respectively.
The top and bottom dashed lines show the
negative values of $P_{\rm diff}$, which
would be obtained from $P_{\rm diff}^{(1)}$  using the Drude
and plasma models, but with omitted constant term on the
right-hand side of Eq.~(\ref{eq13}). The value of this term
is indicated by the short-dashed line at the top of Fig.~\ref{fg8}.
The sensitivity of the CANNEX test to pressure differences
$P_{\rm diff}$ is twice the one to
pressure, i.e., 2~nPa. This means that
differences between the top and bottom solid (dashed) lines
in Fig.~\ref{fg8}, as well as differences between the solid and
dashed lines, can be easily discriminated by comparing the
measurement results with theory. Because of this, the CANNEX
test should be capable not only to discriminate between two
different approaches to describe the relaxation properties
of conduction electrons in nonequilibrium situations,
but to validate or disprove the presence of separation-independent
terms in the nonequlibrium Casimir pressure as well.

\section{Conclusions and discussion}

In the foregoing we have proposed a novel test on the role of
relaxation properties of free electrons in the
out-of-equilibrium Casimir pressure between two parallel metal-coated plates
kept at different temperatures -- one of which is equal to the ambient
temperature. It is shown that if the metallic coatings are
sufficiently thick, the nonequilibrium  pressures are
determined by the mean of the equilibrium contributions calculated at two
different temperatures and the term independent on separation between
the plates. In this situation the temperature-antisymmetric contribution to
the pressure is equal to zero with a high degree of accuracy. Thus,
the proposed test represents an alternative to the previous suggestion \cite{42},
which is directed to testing the role of relaxation properties in the
latter, antisymmetric, contribution. In doing so, the other two
contributions to the nonequlibrium pressure considered
by us are screened out in Ref. \cite{42}.

To demonstrate the role of the relaxation properties of conduction
electrons in a nonequilibrium situation, computations of the Casimir
pressure as a function of separation were performed for two parallel Au
plates of finite thickness, where the
upper plate is kept at ambient temperature
$T_1=300~$K, and the} temperature of the lower plate is $T_2=500~$K. The
ratio of the Casimir pressures for thermal nonequilibrium and equilibrium was
also investigated as a function of temperature of the lower plate
varying from 300 to 500~K. In all cases computations have been made
by using the extrapolations of the optical data of Au to low
frequencies by means of both the plasma and Drude models. It was
shown that the use of different extrapolations leads to markedly
different theoretical predictions.

Furthermore, the experimental configuration of the CANNEX test,
originally intended to measure the Casimir pressure and pressure
gradient in the plane-parallel geometry in thermal equilibrium,
was modified to allow for different temperatures on the two plates
while preserving high experimental sensitivities. The nonequilibrium
pressure, pressure gradient and contributions to these
quantities due to different temperatures of the lower plate were
computed in the experimental configuration using both the plasma
and Drude models for extrapolations of the optical data to low
frequencies. It was shown that even with a rather small difference of 10~K
between the temperatures of the upper and lower plates, theoretical
predictions for the total nonequilibrium pressure and
pressure gradient, as well as for the terms independent on separation
and contributions due to different temperatures, computed using
the plasma and Drude models, can be reliably discriminated taking
into account the experimental sensitivities.

Thus, the modified CANNEX test could be helpful in the resolution of
the Casimir puzzle actively discussed in the literature for the last two
decades. The situation in this problem is really challenging.
The expression for the Casimir free energy was carefully derived from
first principles in case of dissipation using different theoretical
approaches including the fluctuation-dissipation theorem
\cite{12,53,54,55,56,57,58}. This means that it should be valid also in the
case of the Drude model. In spite of this, in many direct measurements
of the Casimir force and its gradient performed starting in 2003
(see Refs.~\cite{4,13} for a review and more modern experiments \cite{14,15,16,17})
the predictions of the Lifshitz theory with taken into account dissipation of free
electrons were excluded at up to 99\% confidence level.
The predictions of the same theory with omitted dissipation of free electrons were
confirmed. In these experiments, the measurement errors were equal to
a fraction of a percent to compare with the difference between two theoretical
predictions up to 5\%. Based on this, the possible role of some unaccounted
systematic effects was underlined by many authors. The situation has been
changed after the proposed differential measurement scheme \cite{59}
where the predictions of the Lifshitz theory combined with the Drude and plasma
models differ by up to a factor of 1000. After performing
the respective experiment
\cite{18}, the Lifshitz formula combined with the Drude model was excluded with
absolute certainty in spite of the fact that it seems to be well justified
theoretically. The predictions of the Lifshitz theory combined with the plasma
model were again found in good agreement with the measurement data.

A commonly accepted understanding of the roots of the problem is still missing.
It is the authors' opinion that they might go back to the foundations of
quantum statistical physics. According to one of the postulates, the responses
of a physical system to a real electromagnetic field possessing a nonzero strength and
to a fluctuating field characterized by a zero strength but nonzero dispersion
are similar. The dielectric response to a real field can be directly measured and
is described by the Drude model as it is confirmed by abundant evidence.
The dielectric response to a fluctuating field, however, can be observed only
indirectly in phenomena such as the Casimir effect. Thus, the above mentioned
postulate may be treated as  a far reaching extrapolation which requires a
reconsideration basing on the experimental results on measuring the Casimir
interaction.

To conclude, the CANNEX test, originally
proposed to measure the pressure and pressure gradient between parallel
flat plates in equilibrium, may
provide important additional information regarding the role of
relaxation properties of conduction electrons in the
out-of-equilibrium Casimir effect.

\section*{Acknowledgments}
The work of V.M.M. was partially funded by the Russian Foundation
for Basic Research, grant number 19-02-00453 A.
V.M.M.\ was also partially supported by the Russian Government Program
of Competitive Growth of Kazan Federal University.
R.I.P.S.\ was supported by the TU Wien.

%%%%%%%%%%%%%%%%%%%%%%%%%%%%%%%%%%%%%%%%%%%%%%%%%%%%%%%%
%\end{document}
\newpage
%%%%%%%__FIGURE__1__%%%%%%%%%%%%%%%%%%%%
\begin{figure}[b]
\vspace*{-6cm}
\centerline{\hspace*{2.5cm}
\includegraphics{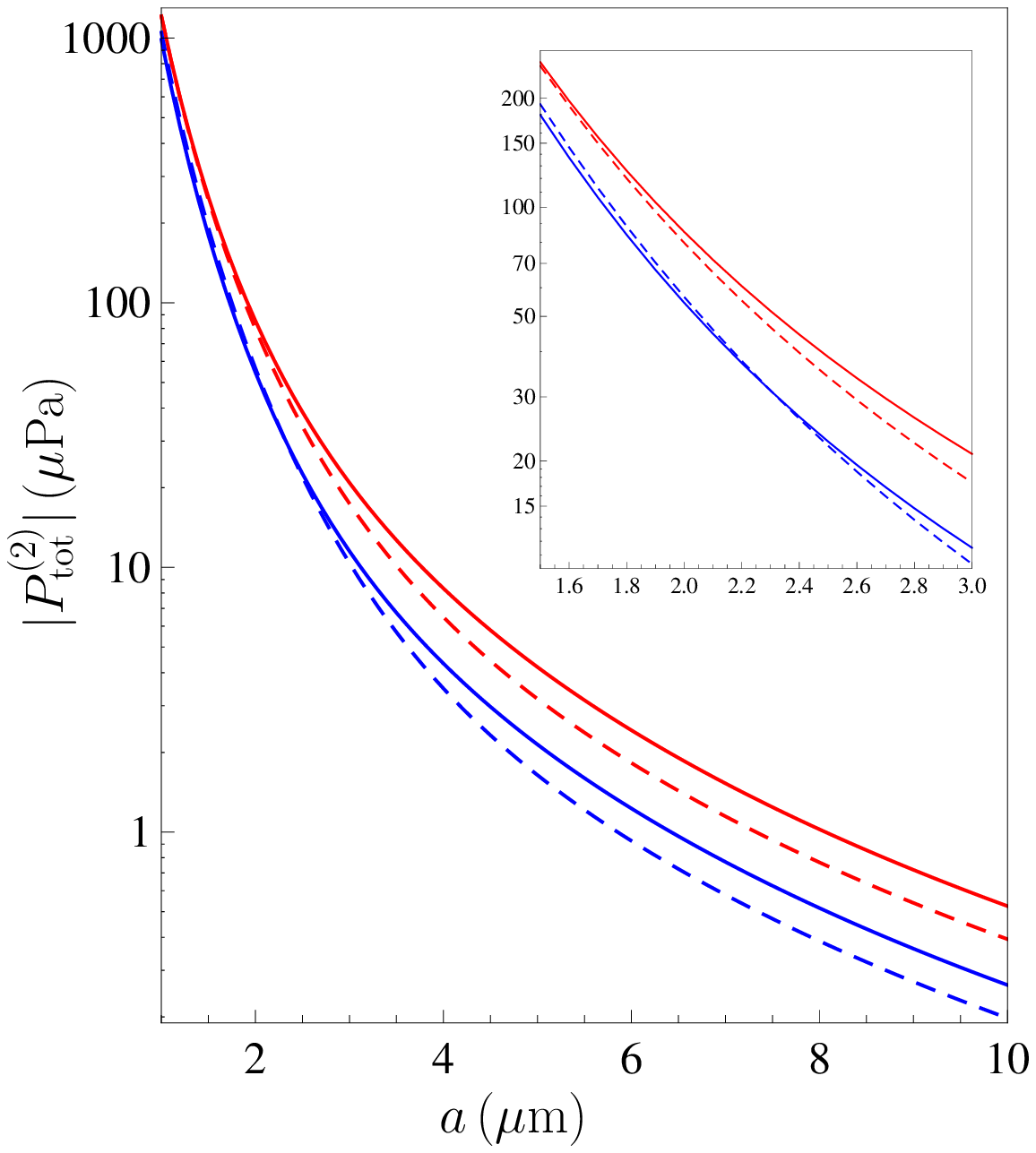}
}
\vspace*{-9.5cm}
\caption{\label{fg1}
The magnitude of the total (Casimir) pressure on the lower Au plate
kept at $T_2=500~$K is shown as a function of separation.
The upper plate is kept at the temperature of the environment $T_1=300~$K.
The top and bottom solid lines are computed using extrapolations of the
optical data to low frequencies by means of the plasma and Drude models,
respectively. The top and bottom dashed lines present the same quantity
in the same way, but in thermal equilibrium $T_1=T_2=300~$K.
The region near an intersection of the solid and dashed bottom lines is
shown in more detail in an inset.
}
\end{figure}
%%%%%%%%%%%%%
%%%%%%%__FIGURE__2__%%%%%%%%%%%%%%%%%%%%
\begin{figure}[b]
\vspace*{-0cm}
\centerline{\hspace*{2.5cm}
\includegraphics{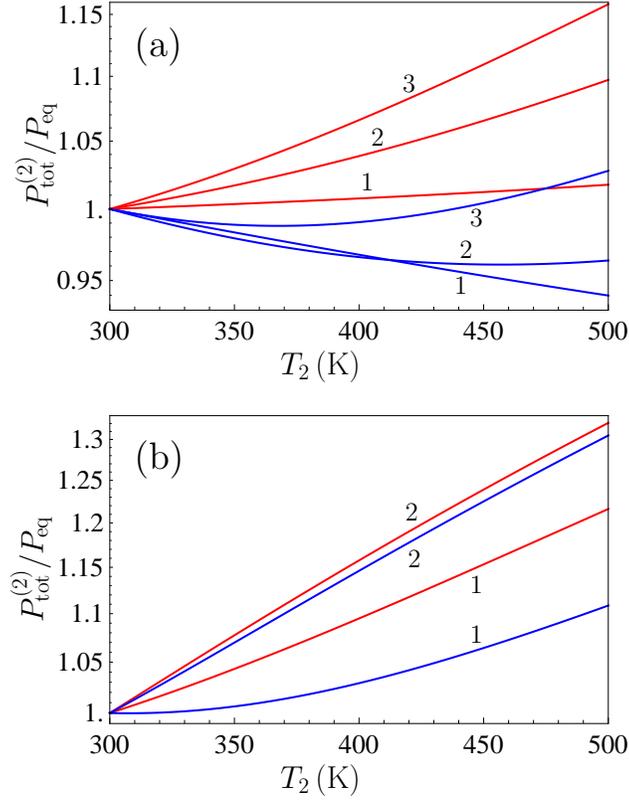}
}
\vspace*{-17.5cm}
\caption{\label{fg2}
The total (Casimir) pressure on the lower Au plate, where the upper plate is
kept at the temperature of the environment $T_1=300~$K, normalized by the
Casimir pressure at thermal
equilibrium, is shown by the pairs of lines labeled (a) 1, 2, and 3 and
(b) 1 and 2 as a function of temperature $T_2$ of the lower plate.
The pairs of lines are computed at separations between the plates
(a) $a=1$, 2, and $2.5~\mu$m for the pairs labeled 1, 2, and 3 and
(b) $a=3$ and $5~\mu$m for the pairs labeled 1 and 2, respectively.
In each pair the top and bottom lines are computed using extrapolations of
the optical data to low frequencies by means of the plasma and Drude models,
respectively.
}
\end{figure}
%%%%%%%%%%%%%
%%%%%%%__FIGURE__3__%%%%%%%%%%%%%%%%%%%%
\begin{figure}[b]
\vspace*{-7cm}
\centerline{\hspace*{2.5cm}
\includegraphics{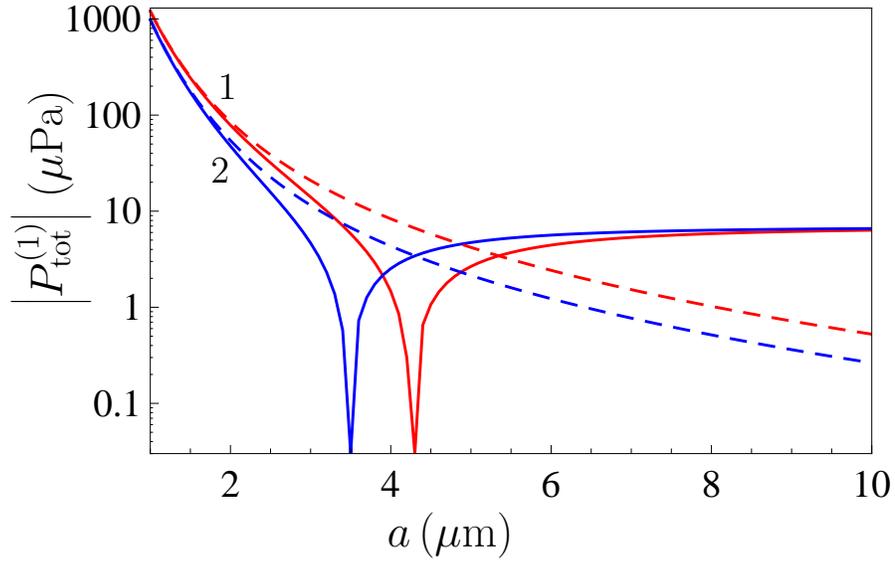}
}
\vspace*{-9.5cm}
\caption{\label{fg3}
The magnitude of the total pressure on the upper plate in the
 configuration of two Au plates
kept at $T_1=300~$K  and $T_2=500~$K is shown as a function of separation
by the two solid lines 1 and 2 computed using extrapolations of the
optical data to low frequencies by means of the plasma and Drude models,
respectively. The dashed lines 1 and 2 show the magnitude of the
Casimir pressure on the lower Au plate computed using the
plasma and Drude models.
}
\end{figure}
%%%%%%%%%%%%%
%%%%%%%__FIGURE__4__%%%%%%%%%%%%%%%%%%%%
\begin{figure}[b]
\vspace*{-17cm}
\centerline{\hspace*{7.5cm}
\includegraphics{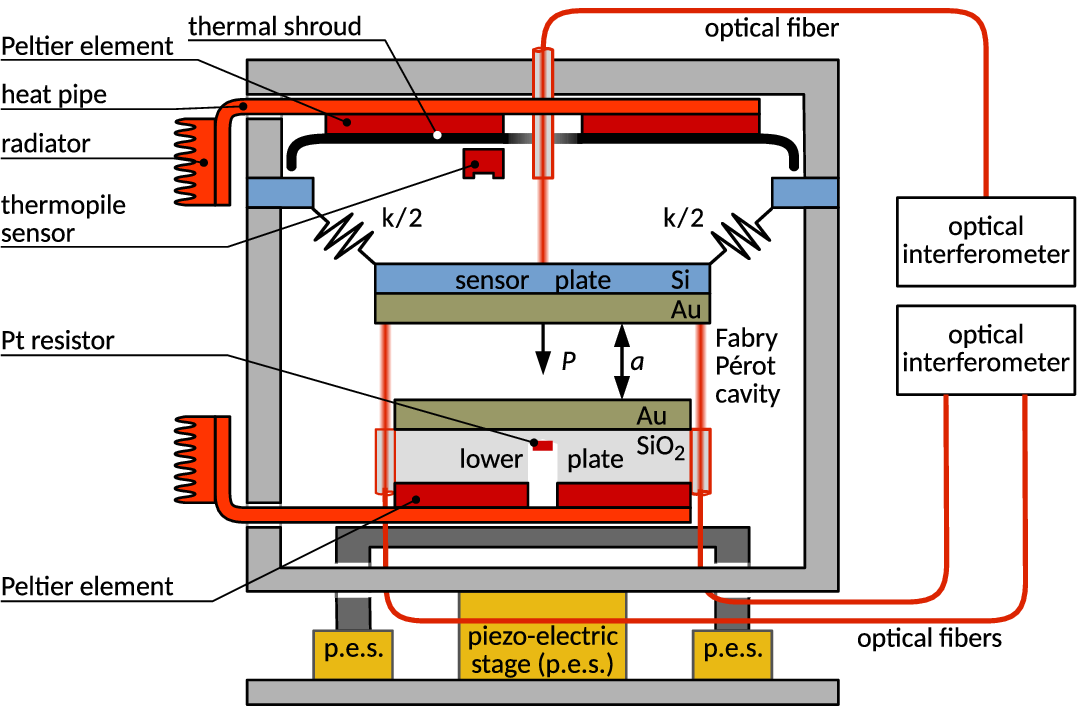}
}
\vspace*{0.5cm}
\caption{\label{fg4}
Simplified schematic of the modified experimental setup of the CANNEX test
that allows for different temperatures on the sensor (upper) and
lower plates by means of accurate feedback-controlled heating and
cooling (see the text for further discussion).
Shown not to scale.
}
\end{figure}
%%%%%%%%%%%%%
%%%%%%%__FIGURE__5__%%%%%%%%%%%%%%%%%%%%
\begin{figure}[b]
\vspace*{-8cm}
\centerline{\hspace*{2.5cm}
\includegraphics{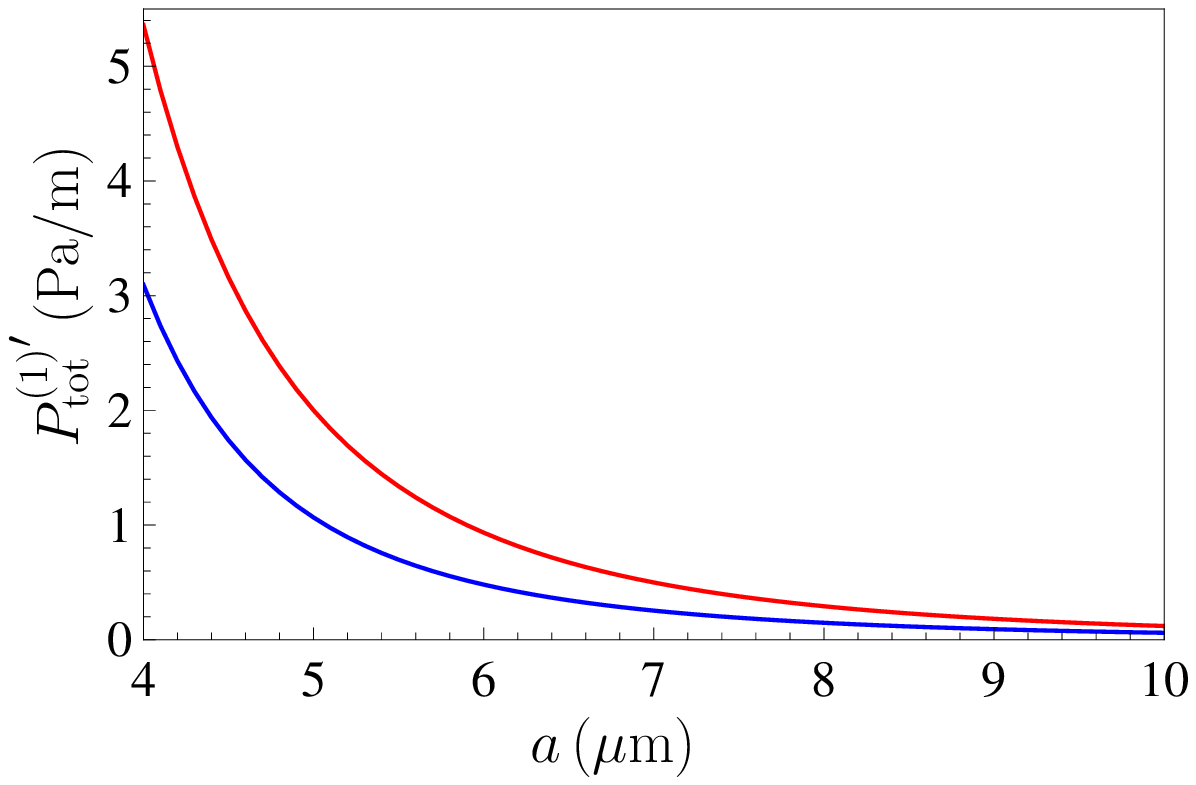}
}
\vspace*{-9.5cm}
\caption{\label{fg5}
The gradient of the total pressure for the experimental parameters of CANNEX
applied to the upper plate kept
at $T_1=300~$K while the lower plate is heated to $T_2=310~$K is shown as a
function of
separation by the top and bottom lines computed using the
extrapolations of the optical data to low frequencies by
means of the plasma and Drude models, respectively.
}
\end{figure}
%%%%%%%%%%%%%
%%%%%%%__FIGURE__6__%%%%%%%%%%%%%%%%%%%%
\begin{figure}[b]
\vspace*{-8cm}
\centerline{\hspace*{2.5cm}
\includegraphics{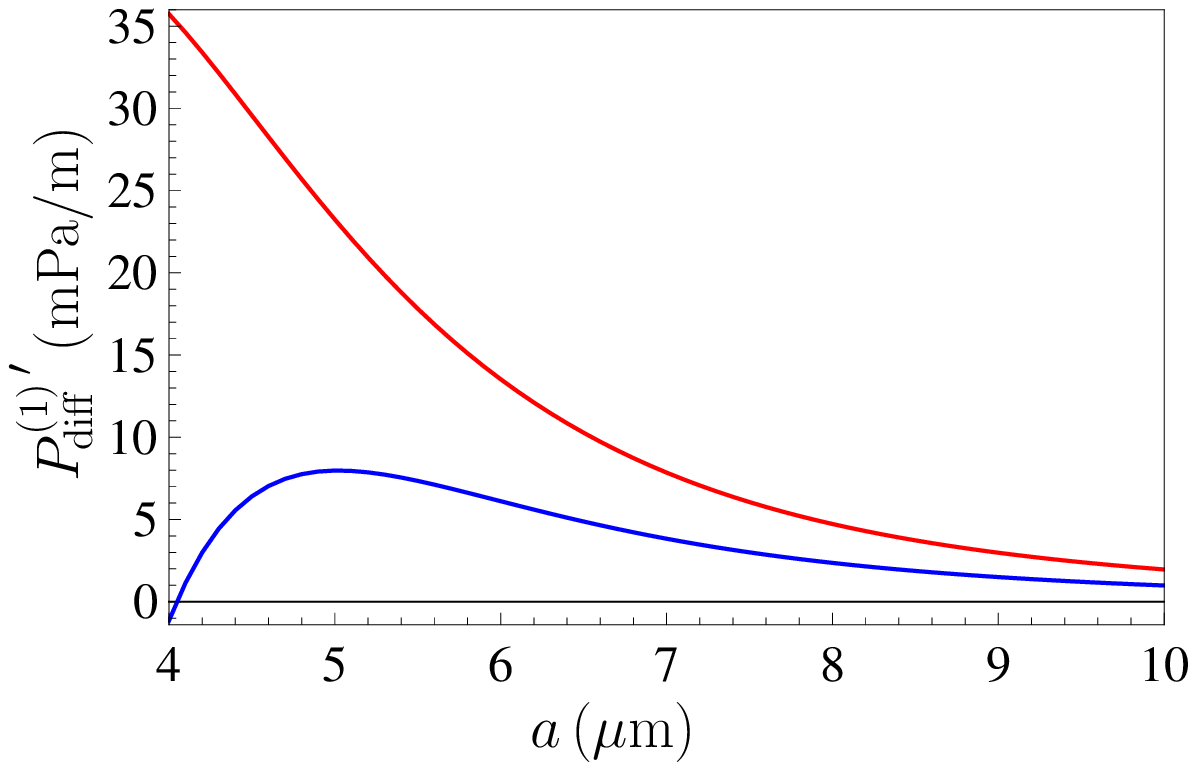}
}
\vspace*{-9.5cm}
\caption{\label{fg6}
The differential gradient of the total pressure
for the experimental parameters of CANNEX
applied to the upper plate kept
at $T_1=300~$K while the lower plate is heated to $T_2=310~$K
is shown as a function of
separation by the top and bottom lines computed using the
extrapolations of the optical data to low frequencies by
means of the plasma and Drude models, respectively.
}
\end{figure}
%%%%%%%%%%%%%
%%%%%%%__FIGURE__7__%%%%%%%%%%%%%%%%%%%%
\begin{figure}[b]
\vspace*{-8cm}
\centerline{\hspace*{2.5cm}
\includegraphics{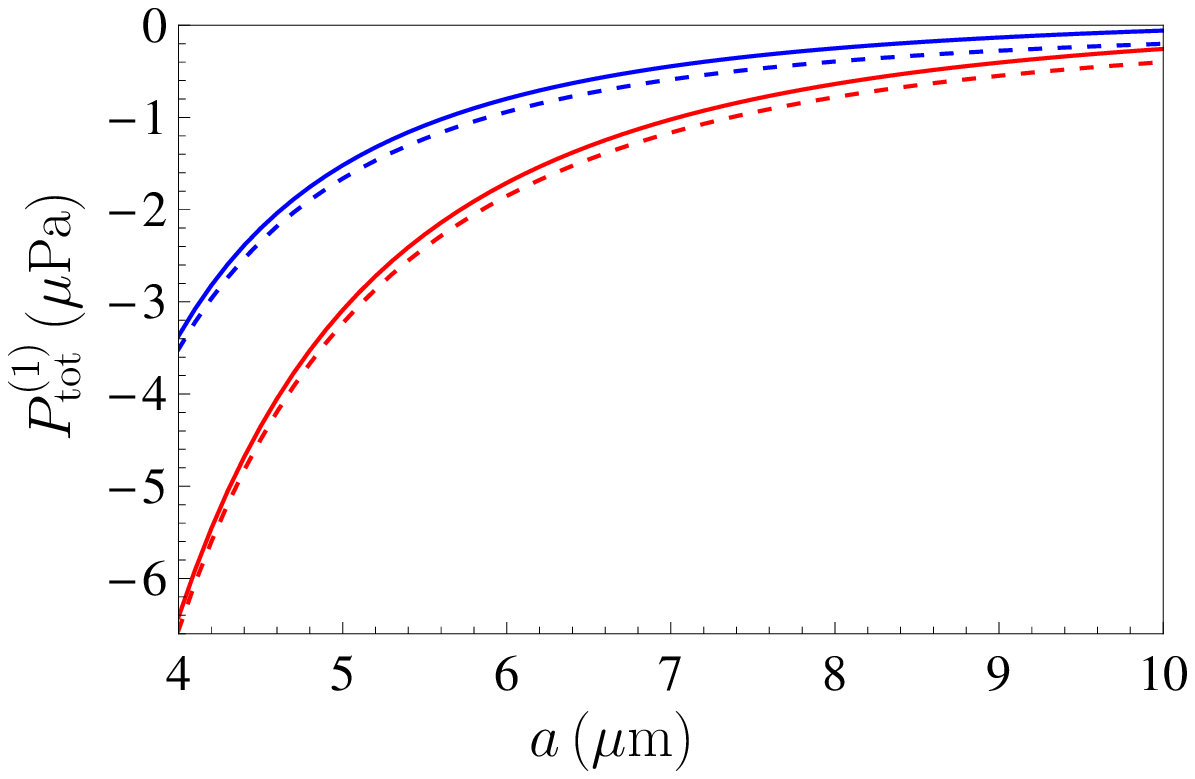}
}
\vspace*{-9.5cm}
\caption{\label{fg7}
The total pressure
for the experimental parameters of CANNEX
applied to the upper plate kept
at $T_1=300~$K while the lower plate is heated to $T_2=310~$K
is shown as a function of
separation by the top and bottom solid lines computed using the
extrapolations of the optical data to low frequencies by
means of the Drude and plasma models, respectively.
Similar results computed with omitted separation-independent term
are shown by the top and bottom dashed lines.
}
\end{figure}
%%%%%%%%%%%%%
%%%%%%%__FIGURE__8__%%%%%%%%%%%%%%%%%%%%
\begin{figure}[b]
\vspace*{-8cm}
\centerline{\hspace*{2.5cm}
\includegraphics{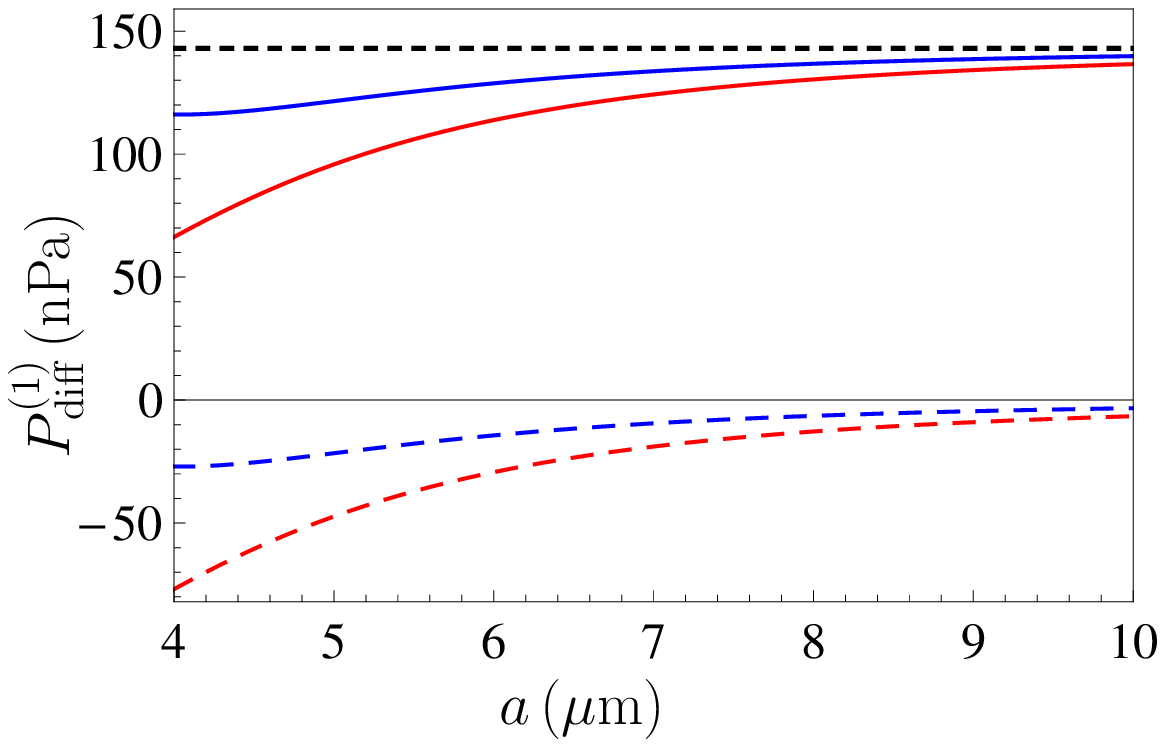}
}
\vspace*{-9.5cm}
\caption{\label{fg8}
The differential  pressure
for the experimental parameters of CANNEX
applied to the upper plate kept
at $T_1=300~$K while the lower plate is heated to $T_2=310~$K
is shown as a function of
separation by the top and bottom solid lines computed using the
extrapolations of the optical data to low frequencies by
means of the Drude and plasma models, respectively.
Similar results computed with the same parameters
but omitted separation-independent term
are shown by the top and bottom
dashed lines for which $P_{\rm diff}<0$.
The separation-independent contribution to the nonequilibrium
pressure is indicated by the short-dashed line.
}
\end{figure}
%%%%%%%%%%%%%
\end{document}